%% file: alicepreprint_CDS.tex
\documentclass[ALICE,manyauthors]{cernphprep}

\usepackage[comma,square,numbers,sort&compress]{natbib}
\usepackage{hyperref}
\usepackage{lineno}
\usepackage{rotating}
\usepackage{graphicx}  
\usepackage{dcolumn}   
\usepackage{bm}        
\usepackage{longtable}
\usepackage{siunitx}
\usepackage{amsmath}
\usepackage{ulem}
\usepackage{epstopdf}
\usepackage{listings}
\usepackage{color}
\usepackage{breakcites}

\input{Commands.tex}

\begin{document}%

\begin{titlepage}
\PHyear{2015}
\PHnumber{268}      
\PHdate{26 September}  
%

\title{Measurement of an excess in the yield of J/$\psi$ at very low $\textbf{\textit{p}}_{\bf T}$ \protect\\ in Pb--Pb collisions at $\mathbf{\sqrt{{\textit s}_{NN}}}$ = 2.76 TeV}
\ShortTitle{Excess in the yield of J/$\psi$ at very low $\textit{p}_{\rm T}$}
\Collaboration{ALICE Collaboration\thanks{See Appendix~\ref{app:collab} for the list of collaboration members}}
\ShortAuthor{ALICE Collaboration} 

\begin{abstract}

\input{ABSTRACT.tex}

\end{abstract}
\end{titlepage}
\setcounter{page}{2}

%
%
%
%

\input{A_Introduction.tex}

\input{B_Experiment.tex}

\input{C_RAA.tex}

\input{D_Excess.tex}

\input{E_Conclusions.tex}

\newenvironment{acknowledgement}{\relax}{\relax}
\begin{acknowledgement}
\section*{Acknowledgements}
\input{acknowledgements_Sept2015.tex}    
\end{acknowledgement}

\bibliographystyle{utphys}   
\bibliography{JpsiLowPtExcessMod}

\newpage
\appendix
\section{The ALICE Collaboration}
\label{app:collab}
\input{Alice_Authorlist_2015-Sep-18.tex}  
\end{document}

%% file: Commands.tex
\def\pt{\mbox{$p_{\rm T} $}} 
   
\def\snn{\mbox{$\sqrt{s_{\rm NN}}$}}   
\def\pb {Pb\mbox{--}Pb }
\def\ppb {p\mbox{--}Pb }

\def\j {\ensuremath{\mathrm{J}\kern-0.02em/\kern-0.05em\psi} }
\def\jwospace {\ensuremath{\mathrm{J}\kern-0.02em/\kern-0.05em\psi}}

\def\raa {\mbox{$R_{\rm{AA}}$}}

\newcommand{ \be }{\begin{eqnarray}}
\newcommand{ \ee }{\end{eqnarray}}

\newcommand {\cent}[2]      {\ensuremath{\mathrm{#1\mbox{--}#2\%}}}
\newcommand {\Npart}      {\ensuremath{\langle N_{\mathrm{part}} \rangle}}
\newcommand {\mevc}      {\ensuremath{\mathrm{MeV}\kern-0.05em/\kern-0.02em c}}
\newcommand {\mevcc}      {\ensuremath{\mathrm{MeV}\kern-0.05em/\kern-0.02em c^2}}
\newcommand {\gevc}      {\ensuremath{\mathrm{GeV}\kern-0.05em/\kern-0.02em c}}
\newcommand {\gevcc}     {\ensuremath{\mathrm{GeV}\kern-0.05em/\kern-0.02em c^2}}
\newcommand {\sqrtS}             {\ensuremath{\sqrt{s}}}
\newcommand {\sqrtSE}[2][TeV]    {$\sqrtS = #2~\mathrm{#1}$} 
\newcommand {\fI}             {\ensuremath{f_\mathrm{I}}}
\newcommand {\fD}             {\ensuremath{f_\mathrm{D}}}

%% file: ABSTRACT.tex

We report on the first measurement of an excess in the yield of \j at very low transverse momentum (\pt~$<$ 0.3 \gevc) in peripheral hadronic \pb collisions at \snn~=~2.76~TeV, performed by ALICE at the CERN LHC.
Remarkably, the measured nuclear modification factor of \j in the rapidity range $2.5<y<4$ reaches about 7 (2) in the \pt~range 0--0.3~\gevc~in the \cent{70}{90} (\cent{50}{70}) centrality class.
The \j production cross section associated with the observed excess is obtained under the hypothesis that coherent photoproduction of \j is the underlying physics mechanism. 
If confirmed, the observation of \j coherent photoproduction in \pb collisions at impact parameters smaller than twice the nuclear radius opens new theoretical and experimental challenges and opportunities. In particular, coherent photoproduction accompanying hadronic collisions may provide insight into the dynamics of photoproduction and nuclear reactions, as well as become a novel probe of the Quark-Gluon Plasma.

%% file: A_Introduction.tex

The aim of experiments with ultra-relativistic heavy-ion collisions is the study of nuclear matter at high temperature and pressure, where Quantum Chromodynamics (QCD) predicts the existence of a deconfined state of hadronic matter, the Quark-Gluon Plasma (QGP). Heavy quarks are expected to be produced in the primary partonic scatterings and to interact with this partonic matter, making them ideal probes of the QGP.  According to the color screening mechanism~\cite{Matsui:1986dk}, quarkonium states are suppressed in the QGP, with different dissociation probabilities for the various states depending on the temperature of the medium. On the other hand, regeneration models predict  charmonium production via the (re)combination of charm quarks during~\cite{Zhao:2011cv,Liu:2009nb,Thews:2000rj} or at the end~\cite{BraunMunzinger:2000px,Andronic:2011yq} of the deconfined phase. 
ALICE measurements of the \j nuclear modification factor (\raa)~\cite{Abelev:2012rv,Abelev:2013ila,Adam:2015rba,Adam:2015isa} and elliptic flow~\cite{ALICE:2013xna} in \pb collisions at a center-of-mass energy of \snn~= 2.76~TeV, as well as the comparison of the \j nuclear modification factor in \ppb collisions at \snn~= 5.02~TeV~\cite{Abelev:2013yxa,Adam:2015iga} with that in Pb--Pb, support the regeneration scenario.

In this letter, we report on the measurement of \j production in hadronic \pb collisions at \snn~= 2.76~TeV at very low \pt~(\pt~$<$ 0.3 \gevc). We find an excess in the yield of \j with respect to expectations from hadroproduction. A plausible explanation is that the excess is caused by coherent photoproduction of J/$\psi$. In this process, quasi-real photons coherently produced by the strong electromagnetic field of one of the lead nuclei interact, also coherently, with the gluon field of the other nucleus, to produce a J/$\psi$. This process proceeds, at leading order in perturbative QCD, through the interchange of two gluons in a singlet color state, probing thus the square of the gluon distribution in the target. The coherence conditions impose a maximum transverse momentum for the produced J/$\psi$ of the order of one over the nuclear radius, so the production occurs at very low \pt. The study of \j photoproduction processes in hadron colliders is known in Ultra-Peripheral collisions (UPC) and several results are already available in this field at RHIC\cite{Afanasiev:2009hy} and at the LHC~\cite{Abelev:2012ba,Abbas:2013oua}. These measurements give insight into the gluon distribution of the incoming Pb nuclei over a broad range of Bjorken-$x$ values, providing information complementary to the study of \j hadroproduction in \ppb and \pb collisions. However, coherent \j photoproduction has never been observed in nuclear collisions with impact parameters smaller than twice the radius of the nuclei. Although the extension to interactions where the nuclei interact hadronically raises several questions, e.g.~how the break-up of the nuclei affects the coherence requirement, we find no other convincing explanation. Assuming, therefore, this mechanism causes the observed excess, we obtain the corresponding cross section in the \cent{30}{50}, \cent{50}{70} and \cent{70}{90} centrality classes.

%% file: B_Experiment.tex

The ALICE detector is described in~\cite{Aamodt:2008zz,Abelev:2014ffa}. 
At forward rapidity  ($2.5 < y < 4$) the production of quarkonium states is measured via their $\mu^{+} \mu^{-}$  decay channel in the muon spectrometer down  to $p_{\rm{T}} = 0$.
The Silicon Pixel Detector (SPD), the scintillator arrays (V0) and the Zero Degree Calorimeters (ZDC) were also used in this analysis.  The SPD is located in the central barrel of ALICE, while the V0 and ZDC are located on both sides of the interaction point. The pseudorapidity coverages of these detectors are $|\eta|<2$ (first SPD layer), $|\eta|<1.4$ (second SPD layer),  $2.8 < \eta < 5.1$ (V0A), $-3.7 < \eta < -1.7$ (V0C) and $|\eta|>8.7$ (ZDC).
The SPD provides the coordinates of the primary interaction vertex. 
The minimum bias trigger (MB) required a signal in the V0 detectors at forward and backward rapidity.
In addition to the MB condition, the dimuon opposite-sign trigger ($\mu\mu$MB), used in this analysis, required at least one pair of opposite-sign track segments detected in the muon spectrometer triggering system, each with a \pt~above the 1 \gevc~threshold of the online trigger algorithm. 
The background induced by the beam and electromagnetic processes was further reduced by the V0 and ZDC timing information and by requiring a minimum energy deposited in the neutron ZDC (ZNA and ZNC)~\cite{ALICE:2012aa}. The energy thresholds were $\sim$ 450 GeV for ZNA and $\sim$ 500 GeV for ZNC and were placed approximately three standard deviations below the energy deposition of a 1.38 TeV neutron.
The data sample used for this analysis amounts to about $17 \times 10^{6}$ $\mu\mu$MB triggered \pb collisions, corresponding to an integrated luminosity $\mathcal{L}_{\rm int} \approx 70$$\mu$b$^{-1}$. 
The centrality determination was based on a fit of the V0 amplitude distribution as described in~\cite{Abelev:2013qoq}. A selection corresponding to the 90\% most central collisions was applied; for these events the MB trigger was fully efficient. In each centrality class, the average number of participant nucleons \Npart~and average value of the nuclear overlap function were derived from a Glauber model calculation~\cite{Miller:2007ri}.

%% file: C_RAA.tex

J/$\psi$ candidates were formed by combining pairs of opposite-sign (OS) tracks reconstructed in the geometrical acceptance of the muon spectrometer and matching a track segment above the 1 \gevc~\pt~threshold in the trigger chambers \cite{Adam:2015isa}. 
\begin{figure}[tbh!p]
\hglue -0.5 true cm
{\centering 
\resizebox*{1\columnwidth}{!}{\includegraphics{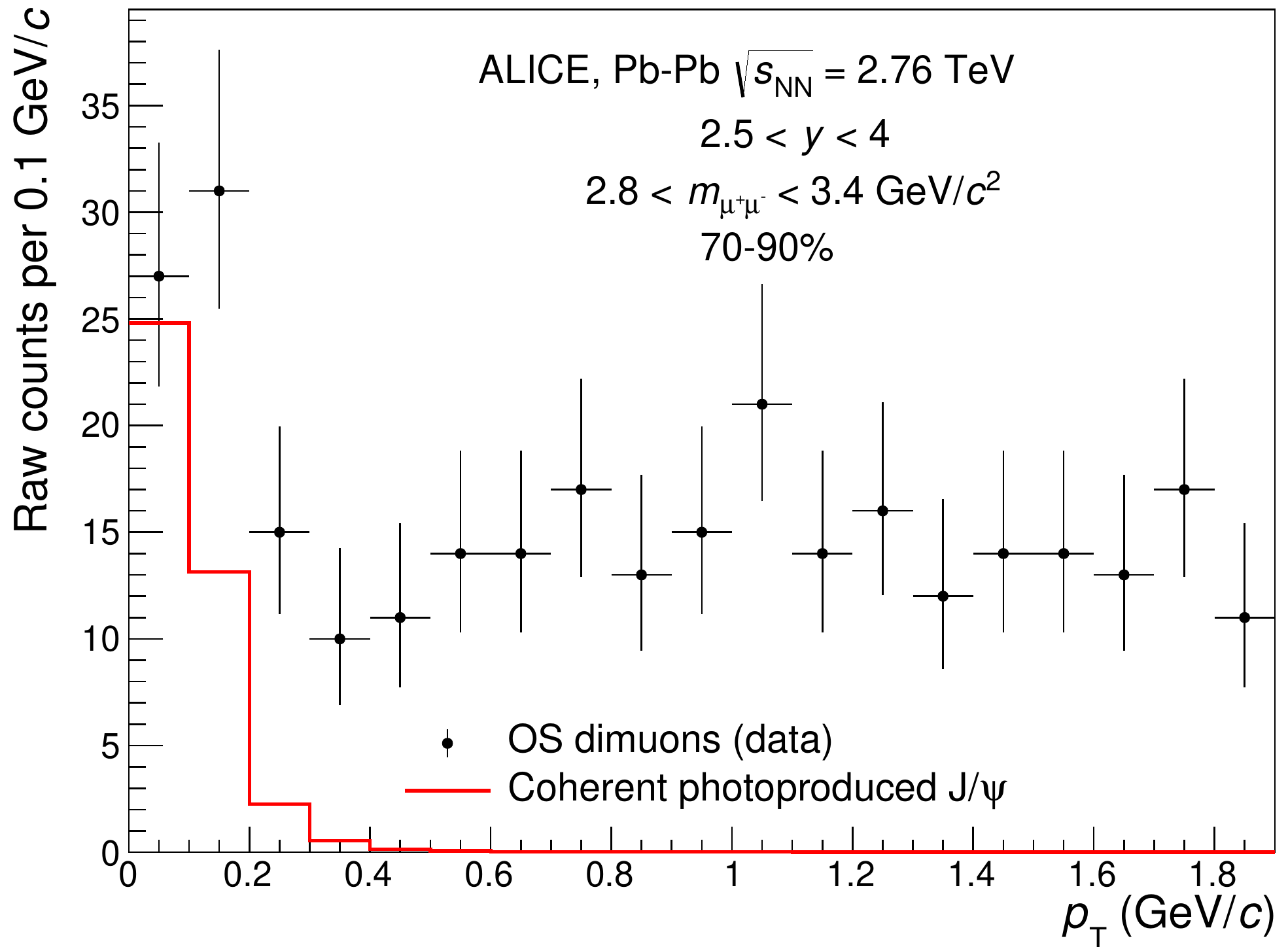}}
\par}
\caption{\label{fig:dimuonpt} (Color online) Raw OS dimuon \pt~distribution for the invariant mass range 2.8~$<$~$m_{\mu^+\mu^-}$~$<$~3.4~\gevcc~and centrality class \cent{70}{90}. Vertical error bars are the statistical uncertainties. The red line represents the \pt~distribution of coherently photoproduced J/$\psi$ as predicted by the STARLIGHT MC generator \cite{starlightMC} in \pb ultra-peripheral collisions and convoluted with the response function of the muon spectrometer. The normalization of the red line is given by the measured number of \j in excess reported in Table~\ref{tab:jpsiexcess} after correction for the $\psi$(2S) feed-down and incoherent contributions (see text).}
\end{figure}
In Fig.~\ref{fig:dimuonpt}, the \pt~distribution of OS dimuons, without combinatorial background subtraction, is shown for the invariant mass range 2.8 $<m_{\mu^+\mu^-}<$ 3.4 \gevcc~in the centrality class \cent{70}{90}. 
A remarkable excess of dimuons is observed at very low \pt~in this centrality class.
 Such an excess has not been observed in the like-sign dimuon \pt~distribution, neither reported in previous measurements in proton-proton collisions~\cite{Abelev:2014qha,Aamodt:2011gj,Abelev:2012rz,Abelev:2012kr,Aaij:2011jh,Aaij:2012asz}. 

\begin{figure}[tbh!p]
{\centering 
\resizebox*{1.\columnwidth}{!}{\includegraphics{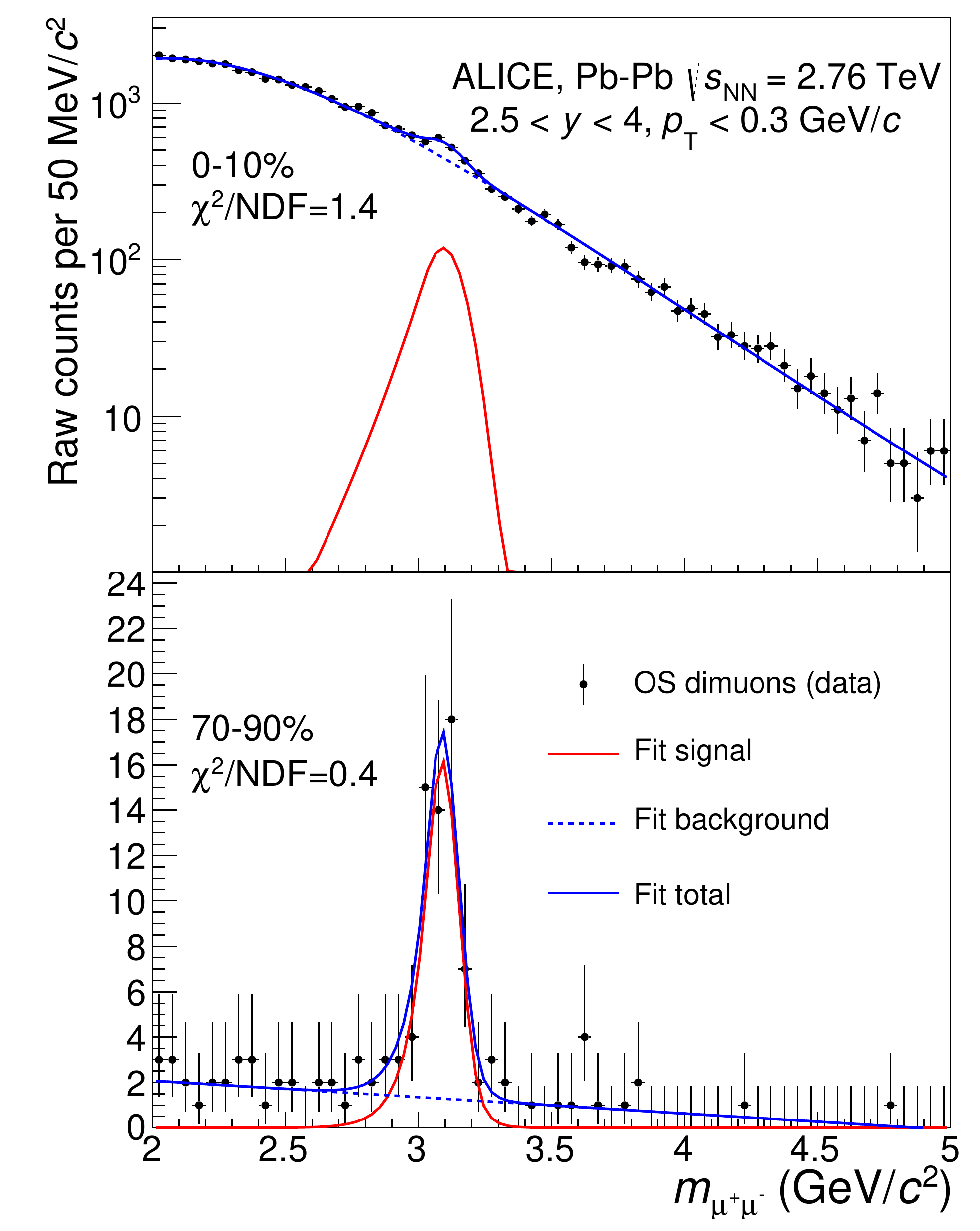}}
\par}
\caption{\label{fig:mass} (Color online) Invariant mass distributions of OS dimuons in the \pt~range 0--0.3 \gevc. The centrality classes are \cent{0}{10} (top) and \cent{70}{90} (bottom). Vertical error bars are the statistical uncertainties.}
\end{figure}

The raw number of \j in five centrality classes (\cent{0}{10}, \cent{10}{30}, \cent{30}{50}, \cent{50}{70} and \cent{70}{90}) and three \pt~ranges (0--0.3, 0.3--1, 1--8 \gevc) was extracted by fitting the OS dimuon invariant mass distribution using a binned likelihood approach. Two functions were considered to describe the \j signal shape: a Crystal Ball function \cite{CBdef} and a pseudo-Gaussian function \cite{Shahoyan}. The tails of the \j signal functions were fixed  using Monte Carlo (MC) simulations for both hadronic~\cite{Abelev:2013ila} and photoproduction hypotheses~\cite{Abelev:2012ba}.
Depending on the \pt~range and centrality class under study, two or three functional forms were used to describe the  background under the \j signal peak. In addition, the fit range was varied. It has also been checked that changing the invariant mass bin width does not significantly modify the results. Fig.~\ref{fig:mass} shows typical fits in the \pt~range 0--0.3 \gevc~for the \cent{0}{10} and \cent{70}{90} centrality classes. The extracted \j signals are the  average of the results obtained making all the combinations of signal shapes, background shapes and fitting ranges, while the systematic uncertainties are given by the RMS of the results. The extracted \j signals and the corresponding statistical and systematic uncertainties are quoted in the second column of Table~\ref{tab:jpsiexcess} for the very low \pt~range.


In each centrality class and \pt~range, the \raa~was obtained from the measured number of \j ($N^{{\rm J}/\psi}_{{\rm AA}}$) corrected for acceptance and efficiency --$({\cal A} \times \epsilon)^{{\rm h~J}/\psi}_{{\rm AA}}$ -- (assuming pure hadroproduction with no polarization), branching ratio ($\rm BR_{\j \rightarrow l^{+} l^{-}}$) and normalized to the equivalent number of MB events ($N_{\rm events}$), average nuclear overlap function ($\left\langle \rm T_{AA} \right\rangle$) and proton-proton inclusive \j production cross section ($\sigma^{{\rm h~J}/\psi}_{{\rm pp}}$), as detailed in \cite{Abelev:2013ila} and shown in Eq.~(\ref{eq:defraa}):  
\begin{equation}
R_{{\rm AA}}^{{\rm h~J}/\psi} = \frac{N^{{\rm J}/\psi}_{{\rm AA}}}{\rm BR_{\j \rightarrow l^{+} l^{-}} \times \it{N}_{\rm events} \times ({\cal A} \times \epsilon)^{{\rm h~J}/\psi}_{{\rm AA}} \times \left\langle \rm T_{AA} \right\rangle \times \sigma^{{\rm h~J}/\psi}_{{\rm pp}}}.
\label{eq:defraa}
\end{equation}

In the \pt~range 1--8 \gevc, the \j cross section in pp collisions at \sqrtSE{2.76}~was directly extracted from the ALICE measurement \cite{Abelev:2012kr}, while in the \pt~ranges 0--0.3 and 0.3--1 \gevc, due to limited statistics, it was obtained by fitting the measured \pt~distribution with the following parametrization \cite{Bossu:2011qe}:
\begin{equation}
\frac{{\rm d}^{2}\sigma^{{\rm h~J}/\psi}_{{\rm pp}}}{{\rm d}p_{\rm T}{\rm d}y} =  \frac{c \times \sigma_{\rm \j} \times p_{\rm T}}{1.5 \times \langle p_{\rm T} \rangle^{2}} \left( 1 + a^{2} \left( \frac{p_{\rm T}}{\langle p_{\rm T} \rangle} \right)^{2}  \right)^{-n},
\label{eq:ptdist}
\end{equation}
where $a=\Gamma(3/2)\Gamma(n-3/2)/\Gamma(n-1)$, $c= 2a^{2}(n-1)$, and $\sigma_{\rm J/\psi}$, $\langle p_{\rm T} \rangle$ and $n$ are free parameters of the fit. A L\'evy--Tsallis function \cite{Tsallis:1987eu,Abelev:2006cs} and UA1 function \cite{Albajar:1989an} were also used to fit the data in order to assess systematic uncertainties. In addition, the validity of the procedure was confirmed using the \j data sample in pp collisions at 7 TeV  \cite{Abelev:2014qha}, where the larger statistics at very low \pt~allows for a direct measurement of the cross sections: the values obtained with this procedure in the \pt~ranges 0--0.3 and 0.3--1 \gevc~agree within 11\% (1.2$\sigma$) and 4\% (0.6$\sigma$), respectively, with the measured cross sections.

\begin{figure}[tbh!p]
  {\centering 
\resizebox*{1.\columnwidth}{!}{\includegraphics{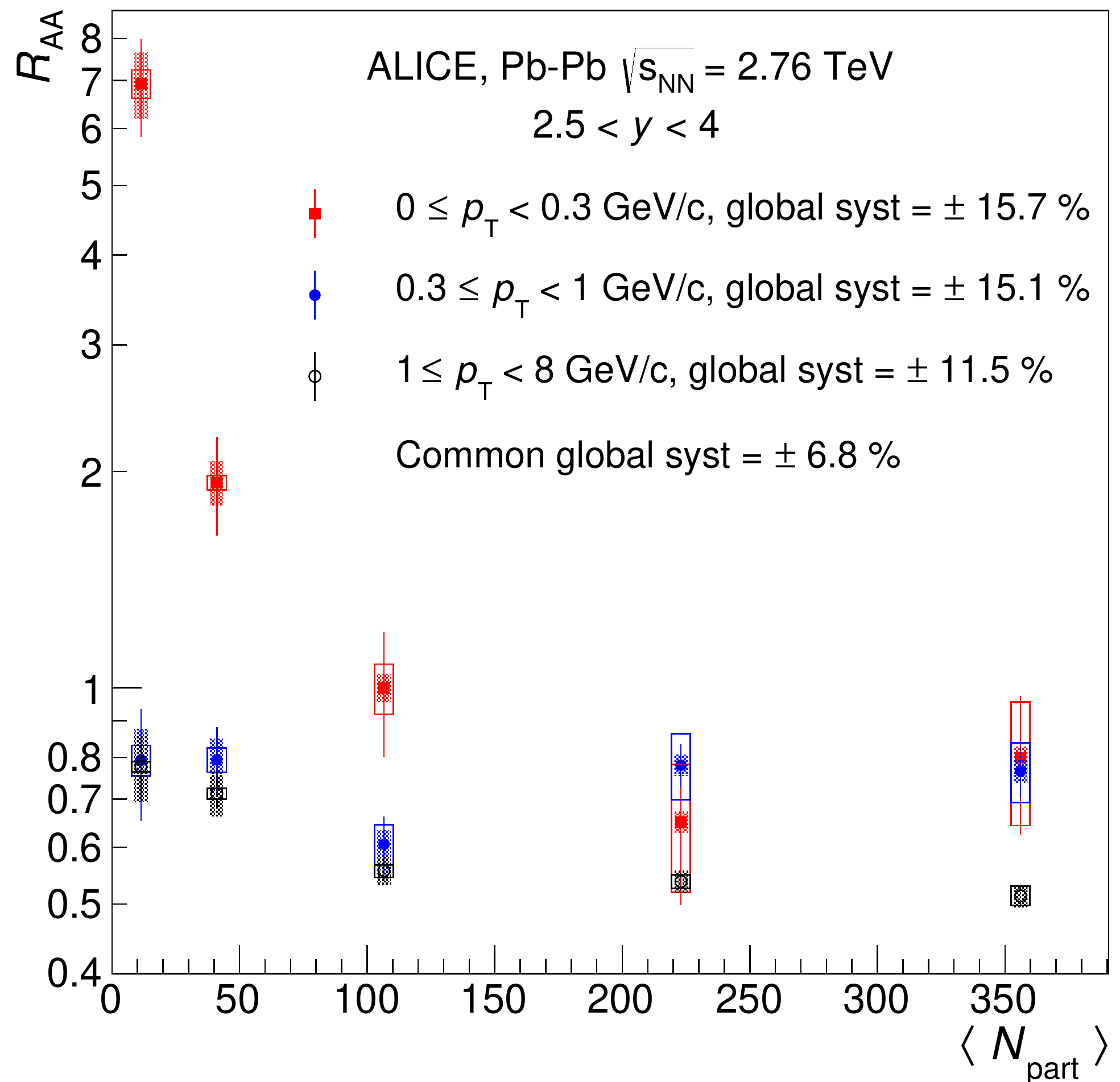}}
\par}
\caption{\label{fig:raa} (Color online) \j \raa~as a function of \Npart~for 3 \pt~ranges in \pb collisions at \snn~= 2.76 TeV. See text for details on uncertainties. When assuming full transverse polarization of the \j in Pb-Pb collisions, as expected if \j are coherently photoproduced, the \raa~values increase by about 21$\%$ in the range 0 $< \pt~ <$ 0.3 GeV/c.}
\end{figure}

The procedures for the determination of the various systematic uncertainties are the same as those followed in~\cite{Abelev:2013ila}, apart from the reference pp cross section in the \pt~ranges 0--0.3 and 0.3--1 \gevc, which incorporate the uncertainties on the fitting procedure described above. In Fig.~\ref{fig:raa}, systematic uncertainties were separated into 4 categories according to their degree of correlation with centrality and \pt: uncorrelated in \pt~and centrality (open boxes), which contain the systematic uncertainties on the signal extraction in \pb (1-23\%); fully correlated as a function of \pt~but not as a function of centrality (shaded areas), which contain the uncertainties on the nuclear overlap function (3.2-7\%), on the determination of the centrality classes (0.7-7.7\%) and on the centrality dependence of the tracking (0-1\%) and trigger efficiencies (0-1\%); fully correlated as a function of centrality but not as a function of \pt~(quoted as global systematics in the legend), which contain the uncertainties on the \j cross section from pp collisions (statistical (3.6-6.9\%) and uncorrelated systematic (3.2-8.0\%)), on the MC input parametrization (0.5-2\%) and on the tracking (10-11\%), trigger (2.2-3.6\%) and matching efficiencies (1\%); and fully correlated in \pt~and centrality (quoted as common global systematics), which contain the correlated systematic uncertainty on the pp reference cross section (5.8\%) and the uncertainty on the number of equivalent minimum bias events (3.5\%).

The \j \raa~shown in Fig.~\ref{fig:raa} exhibits a strong increase in the \pt~range 0--0.3 \gevc~for the most peripheral \pb collisions.  This observation is surprising and none of the transport models~\cite{Zhao:2011cv, Liu:2009nb} that well describe the previous measurements \cite{Abelev:2012rv,Abelev:2013ila,Adam:2015isa} predict such a pattern at LHC energies.

%% file: D_Excess.tex
To quantify the excess of \j at very low \pt, we subtracted the number of \j expected from hadroproduction in \pb collisions.
The following parametrization of the number of hadronic \j ($N^{{\rm h~J}/\psi}_{{\rm AA}}$) as a function of \pt~in a given centrality class was used:
\begin{equation}
\label{eq:expectation}
 \frac{{\rm d}N^{{\rm h~J}/\psi}_{{\rm AA}}}{{\rm d}\pt}  = {\cal N} \times \frac{{\rm d}\sigma^{{\rm h~J}/\psi}_{{\rm pp}}}{{\rm d}\pt} \times R_{{\rm AA}}^{{\rm h~J}/\psi} \times ({\cal A} \times \epsilon)^{{\rm h~J}/\psi}_{{\rm AA}}.
\end{equation}
The factor ${\cal N}$ is fixed by normalizing the integral of Eq.~(\ref{eq:expectation}) in the \pt~range 1--8 \gevc~to the number of \j measured in the same range, where the hadroproduction component is dominant. The second term is given by the fit of the \j \pt-differential cross section measured in pp collisions \cite{Abelev:2012kr} using Eq.~(\ref{eq:ptdist}).
The third term is a parametrization of the $R_{{\rm AA}}^{{\rm h~J}/\psi}$ as a function of \pt~from the ALICE measurements in \pb collisions at 2.76 TeV \cite{Adam:2015isa, Abelev:2013ila}.
These measurements are available in three centrality classes (\cent{0}{20}, \cent{20}{40}, \cent{40}{90}). To calculate the hadroproduction component in the \cent{10}{30} (\cent{30}{50}) centrality class, parameterizations obtained in both \cent{0}{20} and \cent{20}{40} (\cent{20}{40} and \cent{40}{90}) were considered.
A Woods-Saxon like parametrization, which describes the prediction of transport models on \j production in heavy-ion collisions at low $p_{\rm T}$~\cite{Zhao:2011cv, Liu:2009nb}, was used in all the centrality classes:
\begin{equation}
\label{Eq:WoodSaxon}
R_{{\rm AA}}^{{\rm h~J}/\psi}(\pt) = R_{\rm AA}^0 + \frac{\Delta_{R_{\rm AA}}}{1+\exp\Bigl( \frac{p_{\rm T} - p_{\rm T}^0}{\sigma_{p_{\rm T}}} \Bigr) }.
\end{equation}
$R_{\rm AA}^0$, $\sigma_{p_{\rm T}}$ and $\Delta_{R_{\rm AA}}$ are free parameters of the fit while the $p_{\rm T}^{0}$ parameter was either unconstrained or fixed to $M_{\j}$ to force an evolution of $R_{{\rm AA}}^{{\rm h~J}/\psi}$ at very low \pt\ in agreement with the predictions of the transport models~\cite{Zhao:2011cv, Liu:2009nb}. In addition, a first order polynomial and a constant were used in the most peripheral class. Two fitting ranges in \pt\ were considered, either 0-8 or 1-8 \gevc~since the first bin could be biased by the presence of the very low \pt~\j excess.
Finally, the last term in Eq.~(\ref{eq:expectation}) is a parametrization of the acceptance times efficiency of hadronic \j 
($({\cal A} \times \epsilon)^{{\rm h~J}/\psi}_{{\rm AA}}$) -- determined from MC simulations of the muon spectrometer response function -- with either a third-order polynomial or the ratio of two L\'evy--Tsallis functions. Simulations were performed with an embedding technique where MC \j particles are injected into real events and then reconstructed \cite{Abelev:2013ila}. The results of the various parameterizations are averaged in a given range in \pt~and centrality and the RMS of the results is included in the systematic uncertainty on the expected number of hadronic J/$\psi$.

The excess in the number of \j measured in the \pt~range 0--0.3 \gevc~after subtracting the hadronic component is given in the fourth column of Table~\ref{tab:jpsiexcess}.
\begin{table*}[t]
\begin{center}
\begin{tabular}{|c|c|c|c|c|} \hline
Cent. (\%) & $N^{{\rm J}/\psi}_{{\rm AA}}$ & $N^{{\rm h~J}/\psi}_{{\rm AA}}$  & $N^{{\rm excess~J}/\psi}_{{\rm AA}}$  &  d$\sigma_{\j}^{\rm coh}/$d$y$  ($\mu$b)\\ 
\hline
0--10 & $339 \pm 85  \pm 78$ & 406 $\pm$ 14 $\pm$ 55 & $<$ 251  & $<$ 318\\
10--30 & $373 \pm 87 \pm 75$& 397 $\pm$ 10 $\pm$ 61 & $<$ 237 & $<$ 290 \\
30--50 & $187 \pm 37 \pm 15$&  126 $\pm$ 4 $\pm$ 15 & 62 $\pm$ 37 $\pm$ 21 & 73 $\pm$ 44 $^{+26}_{-27}$ $\pm$ 10 \\
50--70 & $89 \pm 13 \pm 2$& 39 $\pm$ 2 $\pm$ 5 & 50 $\pm$ 14 $\pm$ 5 & 58 $\pm$ 16 $^{+8}_{-10}$ $\pm$ 8 \\
70--90 & $59 \pm 9 \pm 3$& 8 $\pm$ 1 $\pm$ 1 & 51 $\pm$ 9 $\pm$ 3 & 59 $\pm$ 11 $^{+7}_{-10}$ $\pm$ 8 \\ \hline
\end{tabular} 
\end{center}
\caption{Raw number of \j ($N^{{\rm J}/\psi}_{{\rm AA}}$), expected raw number of hadronic \j  ($N^{{\rm h~J}/\psi}_{{\rm AA}}$) and measured excess in the number of \j ( $N^{{\rm excess~J}/\psi}_{{\rm AA}}$), all three numbers in the \pt\ range (0--0.3) \gevc, and \j coherent photoproduction cross section in \pb collisons at \snn~= 2.76 TeV, with their statistical and uncorrelated systematic uncertainties. A correlated systematic uncertainty also applies to the cross section. In the most central classes, an upper limit (95$\%$ CL) on the \j yield excess and on the cross section is given.}
\label{tab:jpsiexcess}
\end{table*}
The statistical uncertainty is the quadratic sum of the uncertainties on the measured number of \j in the \pt~ranges 0--0.3 and 1-8 \gevc. The latter is used in the normalization factor of Eq.~(\ref{eq:expectation}). The systematic uncertainty is the quadratic sum of the uncertainties on the signal extraction in 0-0.3 \gevc~(see Table~\ref{tab:jpsiexcess}) and on the parametrization of the hadronic component (13.0$\%$, 12.5$\%$ and 12$\%$ in the \cent{70}{90}, \cent{50}{70} and \cent{30}{50} centrality classes, respectively, see Table~\ref{tab:jpsiexcess}). The significance of the excess is 5.4$\sigma$, 3.4$\sigma$ and 1.4$\sigma$ in the \cent{70}{90}, \cent{50}{70} and \cent{30}{50} centrality classes, respectively. For the two central classes, only the 95\% confidence level limit could be computed. To cross-check the robustness of these results, the excess was re-evaluated assuming a rough parametrization of the $R_{{\rm AA}}^{{\rm h~J}/\psi}$ based on two extreme cases: (i) a constant suppression independent of \pt~($R_{{\rm AA}}^{{\rm h~J}/\psi}$(\pt~$<$ 0.3 \gevc) = $R_{{\rm AA}}^{{\rm h~J}/\psi}$(1 $<$ \pt~$<$ 8~\gevc)), which minimizes the hadronic contribution, and (ii) no suppression at all at low \pt~($R_{{\rm AA}}^{{\rm h~J}/\psi}$(\pt~$<$ 0.3~\gevc) = 1), which gives the maximum possible hadronic contribution. Even with these simplified and extreme assumptions, the \j excess remains significant and compatible with the results reported in Table~\ref{tab:jpsiexcess} within less than 1 (3) times the quoted systematic uncertainty for the \cent{70}{90} (\cent{50}{70}) centrality class.

A plausible explanation of the measured excess is \j photoproduction.
The cross section for this process increases with energy and at the LHC becomes comparable to the \j hadronic cross section. Moreover, the shape of the \pt~distribution in the region of the observed excess is similar to that of a coherently photoproduced \j \cite{Abelev:2012ba}, where the photon is emitted by the electromagnetic field of the source nucleus, and then the target nucleus interacts coherently with the photon to produce the \jwospace, like in \pb ultra-peripheral collisions. The average transverse momentum of coherently photoproduced \j is around 0.055 GeV/$c$. Detailed MC simulations show that detector effects widen reconstructed distribution by approximately a factor of two (see red line in Fig.~\ref{fig:dimuonpt}) and that 98$\%$ of coherently photoproduced \j are contained in the \pt~interval [0, 0.3] GeV/$c$.

Assuming that coherent photoproduction causes the excess at very low \pt, the corresponding cross section can be obtained as described in reference \cite{Abelev:2012ba}. 
The fraction of processes where the coherently emitted photon couples only to single nucleon, so called incoherent photoproduction of \jwospace, and passed the data selection is $\fI = 0.14^{+0.16}_{-0.05}$, while the contribution of coherently produced $\psi(2S)$ with a \j among the decay products which passes the data selection is $\fD = 0.10 \pm 0.06$. Both fractions are used to correct the found excess to extract the number of coherent \jwospace.
This number was then corrected for the acceptance times efficiency (${\cal A} \times \epsilon$ = 11.31 $\pm$ 0.04$\%$) taking into account that photoproduced \j are expected to be transversally polarized, for branching ratio and normalized to the integrated luminosity and the width of the rapidity range.
 For the \cent{70}{90} centrality class, the cross section per unit of rapidity amounts to 59 $\pm$ 11 (stat)$^{+7}_{-10}$ (uncor. syst) $\pm$ 8 (cor. syst) $\mu$b (see Table~\ref{tab:jpsiexcess}, where the values for the other centrality classes are also reported). The uncorrelated centrality dependent systematic uncertainties contain, in addition to the one on the measured excess, the uncertainties on the incoherent and $\psi(2S)$ feed-down contributions (see above), on the determination of the centrality classes (0.7-7.7$\%$), on the trigger efficiency (0-1$\%$), on the tracking efficiency (0-1$\%$) and on the tracking and trigger efficiency loss as a function of centrality (0-3$\%$). The correlated systematic uncertainties contain the uncertainty on the branching ratio (1$\%$), on the luminosity ($^{+7.8}_{-6.5}\%$), on the tracking (11$\%$), trigger (3.6$\%$) and matching efficiencies (1$\%$) and on the MC input parametrization (3$\%$).

In UPC of lead nuclei at  \snn~=~2.76~TeV one expects the incoherent yield in the \pt~range 0.3--1~\gevc~to be about 30$\%$ of the coherent yield in the \pt~range 0--0.3 \gevc~\cite{Abelev:2012ba}. Assuming the same behavior in peripheral collisions, one would expect a 23$\%$ (4$\%$) contribution of incoherent \j to the total number of \j measured in the \cent{70}{90} (\cent{50}{70}) centrality class in the \pt~range 0.3--1 \gevc. 
The significance of the present data sample is not sufficient to confirm the presence of incoherent photoproduction in this \pt~range.

The probability of a random coincidence of a MB collision and a coherent production of a \j in a UPC satisfying the dimuon trigger, in the same bunch crossing, has been evaluated. In the overall data sample, only one random coincidence is expected for the full centrality range, corresponding to 0.6 coincidences in the \cent{30}{90} centrality class.

To our knowledge there is no numerical prediction for the cross section of coherent photoproduction of \j in peripheral collisions. Given that the nuclei also undergo a hadronic interaction, it is not clear how to incorporate the coherence conditions. 
To have a rough estimate, we considered the extreme assumption that all the charges in the source and all the nucleons in the target contribute to the photonuclear cross section as in coherent UPC (see also~\cite{Klusek-Gawenda:2015hja}). The photon flux, see e.g.~\cite{Baur:2001jj}, was obtained integrating in the impact parameter range corresponding to the centrality class. We used two different approaches: the vector dominance model of~\cite{Klein:1999qj}, normalized to the measured UPC data~\cite{Abelev:2012ba,Abbas:2013oua}, and the perturbative QCD model of~\cite{Baur:2001jj} with the parameterization of~\cite{Guzey:2013xba}. In both cases we obtain a cross section in the \cent{70}{90} centrality class of about 40 $\mu$b, which is of the same order of magnitude as our measurement.
Note that the most peripheral class corresponds to the hadronic interaction of just a few nucleons ($N_{\rm part}\approx 11$), so the interaction is close to the ultra-peripheral case and the comparison to the estimate seems reasonable. Another interesting hypothesis, not considered, would be that only the spectators in the target are the ones that interact coherently with the photon. In this case, the \pt~distribution of the excess would get wider as the centrality increases, providing an experimental tool to discriminate among potential models. Indeed, as the size of the spectator region decreases with centrality, the maximum \pt, given by the coherence condition and the uncertainty principle, would increase.

%% file: E_Conclusions.tex

In summary, we reported on the ALICE measurement of \j production~at very low \pt~and forward rapidity in \pb collisions at \snn~= 2.76 TeV. 
A strong increase of the \j \raa~is observed in the range $0 \leq p_{\rm T}<0.3$~\gevc~for the \cent{70}{90} (\cent{50}{70}) centrality class, where \raa~reaches a value of about 7 (2).
The excess has been quantified with a significance of 5.4 (3.4) $\sigma$ assuming a smooth evolution of the \j hadroproduction at low \pt. 
Coherent photoproduction of \j is a plausible physics mechanism at the origin of this excess. 
Following this assumption, the coherent photoproduction cross section has been extracted for the centrality classes \cent{30}{50}, \cent{50}{70} and \cent{70}{90}  while an upper limit is given for \cent{0}{10} and \cent{10}{30}.  
It would be very challenging for existing theoretical models, which only include hadronic processes, to explain this excess.  The survival of an electromagnetically produced charmonium in a nuclear collision merits  theoretical investigation.
In addition, coherent photoproduced \j may be formed in the initial stage of the collisions and could therefore interact with the QGP, resulting in a modification of the measured cross section with respect to the expectation of theoretical models. 
In particular, one expects a partial suppression of photoproduced \j due to color screening of the heavy quark potential in the QGP. The regenerated \j in the QGP exhibit a wider \pt~distribution and do not contribute to the measured excess, making this measurement a potentially powerful tool to constrain the suppression/regeneration components in the models.
Experimentally, the increase of the LHC heavy ion luminosity during Run~2 will lead to a factor 10 larger data sample, thus improving the precision of the present measurement and  opening the possibility to determine whether  the \j excess at very low \pt~is also present in the most central collisions.

%% file: acknowledgements_Sept2015.tex

The ALICE Collaboration would like to thank all its engineers and technicians for their invaluable contributions to the construction of the experiment and the CERN accelerator teams for the outstanding performance of the LHC complex.
The ALICE Collaboration gratefully acknowledges the resources and support provided by all Grid centres and the Worldwide LHC Computing Grid (WLCG) collaboration.
The ALICE Collaboration acknowledges the following funding agencies for their support in building and
running the ALICE detector:
State Committee of Science,  World Federation of Scientists (WFS)
and Swiss Fonds Kidagan, Armenia;
Conselho Nacional de Desenvolvimento Cient\'{\i}fico e Tecnol\'{o}gico (CNPq), Financiadora de Estudos e Projetos (FINEP),
Funda\c{c}\~{a}o de Amparo \`{a} Pesquisa do Estado de S\~{a}o Paulo (FAPESP);
National Natural Science Foundation of China (NSFC), the Chinese Ministry of Education (CMOE)
and the Ministry of Science and Technology of China (MSTC);
Ministry of Education and Youth of the Czech Republic;
Danish Natural Science Research Council, the Carlsberg Foundation and the Danish National Research Foundation;
The European Research Council under the European Community's Seventh Framework Programme;
Helsinki Institute of Physics and the Academy of Finland;
French CNRS-IN2P3, the `Region Pays de Loire', `Region Alsace', `Region Auvergne' and CEA, France;
German Bundesministerium fur Bildung, Wissenschaft, Forschung und Technologie (BMBF) and the Helmholtz Association;
General Secretariat for Research and Technology, Ministry of Development, Greece;
Hungarian Orszagos Tudomanyos Kutatasi Alappgrammok (OTKA) and National Office for Research and Technology (NKTH);
Department of Atomic Energy and Department of Science and Technology of the Government of India;
Istituto Nazionale di Fisica Nucleare (INFN) and Centro Fermi -
Museo Storico della Fisica e Centro Studi e Ricerche ``Enrico Fermi'', Italy;
MEXT Grant-in-Aid for Specially Promoted Research, Ja\-pan;
Joint Institute for Nuclear Research, Dubna;
National Research Foundation of Korea (NRF);
Consejo Nacional de Cienca y Tecnologia (CONACYT), Direccion General de Asuntos del Personal Academico(DGAPA), M\'{e}xico, Amerique Latine Formation academique - European Commission~(ALFA-EC) and the EPLANET Program~(European Particle Physics Latin American Network);
Stichting voor Fundamenteel Onderzoek der Materie (FOM) and the Nederlandse Organisatie voor Wetenschappelijk Onderzoek (NWO), Netherlands;
Research Council of Norway (NFR);
National Science Centre, Poland;
Ministry of National Education/Institute for Atomic Physics and National Council of Scientific Research in Higher Education~(CNCSI-UEFISCDI), Romania;
Ministry of Education and Science of Russian Federation, Russian
Academy of Sciences, Russian Federal Agency of Atomic Energy,
Russian Federal Agency for Science and Innovations and The Russian
Foundation for Basic Research;
Ministry of Education of Slovakia;
Department of Science and Technology, South Africa;
Centro de Investigaciones Energeticas, Medioambientales y Tecnologicas (CIEMAT), E-Infrastructure shared between Europe and Latin America (EELA), Ministerio de Econom\'{i}a y Competitividad (MINECO) of Spain, Xunta de Galicia (Conseller\'{\i}a de Educaci\'{o}n),
Centro de Aplicaciones Tecnológicas y Desarrollo Nuclear (CEA\-DEN), Cubaenerg\'{\i}a, Cuba, and IAEA (International Atomic Energy Agency);
Swedish Research Council (VR) and Knut $\&$ Alice Wallenberg
Foundation (KAW);
Ukraine Ministry of Education and Science;
United Kingdom Science and Technology Facilities Council (STFC);
The United States Department of Energy, the United States National
Science Foundation, the State of Texas, and the State of Ohio;
Ministry of Science, Education and Sports of Croatia and  Unity through Knowledge Fund, Croatia;
Council of Scientific and Industrial Research (CSIR), New Delhi, India;
Pontificia Universidad Cat\'{o}lica del Per\'{u}.

%% file: Alice_Authorlist_2015-Sep-18.tex

\begingroup
\small
\begin{flushleft}
J.~Adam$^{\rm 40}$, 
D.~Adamov\'{a}$^{\rm 84}$, 
M.M.~Aggarwal$^{\rm 88}$, 
G.~Aglieri Rinella$^{\rm 36}$, 
M.~Agnello$^{\rm 110}$, 
N.~Agrawal$^{\rm 48}$, 
Z.~Ahammed$^{\rm 132}$, 
S.U.~Ahn$^{\rm 68}$, 
S.~Aiola$^{\rm 136}$, 
A.~Akindinov$^{\rm 58}$
% ...

\bigskip 

\textbf{\Large Affiliation Notes} 

% ...
$^{\rm IV}$ Also at: M.V. Lomonosov Moscow State University, D.V. Skobeltsyn Institute of Nuclear, Physics, Moscow, Russia

\bigskip

\textbf{\Large Collaboration Institutes} 

% ...
$^{36}$ European Organization for Nuclear Research (CERN), Geneva, Switzerland\\
% ...
$^{40}$ Faculty of Nuclear Sciences and Physical Engineering, Czech Technical University in Prague, Prague, Czech Republic\\
% ...
$^{48}$ Indian Institute of Technology Bombay (IIT), Mumbai, India\\
% ...
$^{58}$ Institute for Theoretical and Experimental Physics, Moscow, Russia\\
% ...
$^{68}$ Korea Institute of Science and Technology Information, Daejeon, South Korea\\
% ...
$^{84}$ Nuclear Physics Institute, Academy of Sciences of the Czech Republic, \v{R}e\v{z} u Prahy, Czech Republic\\
% ...
$^{88}$ Physics Department, Panjab University, Chandigarh, India\\
% ...
$^{110}$ Sezione INFN, Turin, Italy\\
% ...
$^{132}$ Variable Energy Cyclotron Centre, Kolkata, India\\
% ...
$^{136}$ Yale University, New Haven, Connecticut, United States\\
% ...
\endgroup